# Off the Grid… and Back Again? The Recent Evolution of American Street Network Planning and Design

Geoff Boeing

Department of Urban Planning and Spatial Analysis
Sol Price School of Public Policy
University of Southern California

**Abstract**

This morphological study identifies and measures recent nationwide trends in American street network design. Historically, orthogonal street grids provided the interconnectivity and density that researchers identify as important factors for reducing vehicular travel and emissions and increasing road safety and physical activity. During the 20$^{th}$ century, griddedness declined in planning practice alongside declines in urban form compactness, density, and connectivity as urbanization sprawled around automobile dependence. But less is known about comprehensive empirical trends across US neighborhoods, especially in recent years. This study uses public and open data to examine tract-level street networks across the entire US. It develops theoretical and measurement frameworks for a quality of street networks defined here as griddedness. It measures how griddedness, orientation order, straightness, 4-way intersections, and intersection density declined from 1940 through the 1990s while dead-ends and block lengths increased. However, since 2000, these trends have rebounded, shifting back toward historical design patterns. Yet, despite this rebound, when controlling for topography and built environment factors all decades post-1939 are associated with lower griddedness than pre-1940. Higher griddedness is associated with less car ownership—which itself has a well-established relationship with vehicle kilometers traveled and greenhouse gas emissions—while controlling for density, home and household size, income, jobs proximity, street network grain, and local topography. Interconnected grid-like street networks offer practitioners an important tool for curbing car dependence and emissions. Once established, street patterns determine urban spatial structure for centuries, so proactive planning is essential.

# 1. Introduction

> "…they spent two hours in this strikingly American town [Salt Lake City], built on the pattern of other cities of the Union, like a checker-board, 'with the sombre sadness of right-angles,' as Victor Hugo expresses it. The founder of the City of the Saints could not escape from the taste for symmetry which distinguishes the Anglo-Saxons." –Jules Verne, *Around the World in Eighty Days*

The orthogonal grid was the primary mode of geometric spatial ordering in US cities from the 18th century through the early 20th century, with only occasional exceptions, such as the picturesque movement in suburban design (Southworth and Ben-Joseph, 1997). But during the 20th century, new automobile-centric transportation technologies and engineering standards emerged to organize new cities and suburbs according to radically different spatial logics (Jackson, 1985; Wheeler, 2008). Around World War II, the automobile's popularity—and urban planning's responses to it—had gradually reached a tipping point. Planners and engineers tried to accommodate shifting cultural preferences and mobility patterns through a new street network design paradigm predicated on winding loops, culs-de-sac, and automobile-oriented suburbanization (Hayden, 2004; Southworth and Ben-Joseph, 1995). As griddedness, connectivity, and density declined, block sizes and street circuity grew. The resulting sprawl shapes American life today (Forsyth and Southworth, 2008).

Hundreds of studies in recent decades have identified the role that street network design plays in travel behavior, public health, and environmental sustainability. Traditional patterns, such as fine-grained interconnected grids, are associated with higher rates of active transportation and less driving. But after a century of building cities around the spatial logic of the automobile, planners today face car-oriented crises in public safety, physical inactivity, traffic congestion, and rising environmental pollution and greenhouse gas (GHG) emissions. This is of enormous importance to planning practice—which sits at a critical leverage point to shape these outcomes. Over the past 25 years, planning scholars and prominent practitioner groups such as the Congress for the New Urbanism (CNU) have highlighted the links between physical planning and these transportation and environmental crises. How has US planning practice responded to these calls?

This study offers a comprehensive empirical analysis of all US streets to analyze these trends at neighborhood scales. It explores how US street network design has changed over time, especially in recent years, and considers what this means for automobility and its second-order effects. It makes two primary contributions. First, it measures exactly how street network design grew more coarse-grained, disconnected, and circuitous nationwide over the 20th century before rebounding over the past 20 years, offering a new scorecard to assess planning practice's progress toward public health and sustainability goals. Second, it identifies new relationships between vintage, urban form, and car dependence while employing previously underutilized but essential topographical controls. To accomplish this, it develops a theoretical and measurement framework for street network griddedness as well as several new urban form vintage estimation methods and algorithms.

In sum, this study provides a new quantitative post-war history of American street network geography nationwide at the local scale. First it reviews street patterns in planning history and the literature on network design, sprawl, sustainability, and travel. This literature identifies the advantages of interconnected grids and traces the evolution of network design through primarily historical and case study research, but it tells us less about nationwide empirical patterns and little about recent developments.

This study picks up this research thread to take a deeper look. It uses computational big data methods to model the street networks and vintage of every US census tract, collectively comprising approximately 19 million intersections/dead-ends and 24 million street segments. This offers a new glimpse into the spatial outcomes of planning practice and a critical empirical assessment of its recent directions. In particular, it reflects on what planners can do to continue the promising nascent trend toward more sustainable forms. Retrofitting existing suburbia offers one opportunity, but can be challenging as road networks and land parcels create strong path dependence (Boarnet et al., 2011; Dunham-Jones and Williamson, 2011). But ongoing greenfield development, infill projects, and targeted redevelopment offer practitioners important opportunities to plan proactively for a more sustainable urban future. Street network design cannot merely reflect short-sighted transportation paradigms: due to its near-permanence, planners must carefully plan for decades or centuries of travel behavior, flexibly and sustainably.

## 2. Background

### 2.1. Defining the Grid

Street networks provide a physical substrate and connective tissue that organize a city's human dynamics. A street network's pattern, configuration, and grain reflect prevailing technologies, design paradigms, politics, power, terrain, and local culture and economic conditions (Rose-Redwood and Bigon, 2018). The grid is the world's most ubiquitous planned pattern. From Hippodamus's urban design of Piraeus in ancient Greece, to the Spanish Crown's colonial Law of the Indies, to the New York Commissioners' Plan and Salt Lake City's Plat of Zion, the classic street grid has been used for millennia to impose urbane order over the landscape, to streamline transportation systems, to make land legible to speculation and development, and to organize the city democratically or around symbols of power and spaces of control (Grant, 2001; Groth, 1981; Kostof, 1991; Marcuse, 1987; Martin, 2000; Scheer, 2017).

The word *grid* first appeared in English in 1839 as a back-formation of *gridiron*, a metal grate traditionally used to broil food over an open flame[1]. Its composition of parallel and perpendicular bars has inspired centuries of appropriation for similar geometric patterns, such as the playing field in American football and the design of certain street networks. Much like the gridiron from which its etymology arises, the classic street grid consists of a set of streets characterized by three properties: orientation order, straightness, and four-way junctions. That is, to be a grid, a street network must have an internally-consistent orientation, be relatively straight, and primarily comprise four-way intersections instead of three-way "T" junctions or dead-ends. Each of these components is necessary but alone insufficient for griddedness. Only in unison do they make a street grid, as illustrated by Figure 1.

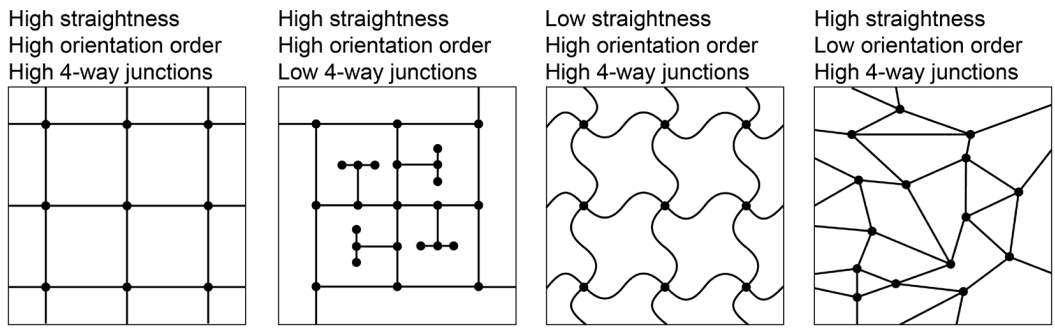

**Figure 1.** Theoretically, a street grid has an internally-consistent orientation, is relatively straight, and comprises mostly four-way intersections. Each of these three characteristics is necessary but alone insufficient: only in unison do they make a street grid.

**2.2. American Street Network Patterns**

The grid has a long history in the US. Some pre-Columbian gridded patterns likely existed, but Spain's 1573 Law of the Indies spread urban grids throughout its American colonies by systematizing the design of rectilinear street networks around central plazas (Low, 2009; Rose-Redwood and Bigon, 2018; Wheeler, 2015). Beyond the Spanish model, other prominent colonial urban grids included Penn's 1682 plan of Philadelphia and Oglethorpe's 1733 plan of Savannah.

Two years after America won its independence, Thomas Jefferson and his Age of Enlightenment associates drafted the Land Ordinance of 1785 followed by the Northwest Ordinances of 1787 and 1789, dividing the American frontier into a regular grid of townships and parcels that shaped subsequent US expansion (Jacobson, 2002). This rationalist-utopian ideal culminated in the US Homestead Act of 1862 which partitioned the Midwest and Great Plains into square miles subdivided into 160-acre quarters to promote rapid westward expansion, settlement, standardization, transportation, legibility, and a sense of civilization in the wilderness (Grant, 2001; Jackson, 1985; Sennett, 1990). To lay out local transportation networks and parcelize land, town planners adopted these preexisting orthogonal spatial frameworks or reoriented them to the local terrain. US town planning subsequently evolved through eras of fine-grained grids, rectangular street-car suburbs, and later degenerate grids (Southworth and Owens, 1993; Wheeler, 2015).

These models of gridded spatial order prevailed into the 20th century, until the close of the 1930s marked a rupture between traditional urbanism and modern automobile dependence. Just a decade earlier, experimental gestures toward the new urban future had appeared: in 1929, Clarence Perry published his neighborhood unit concept—an influential model of physically segregated communities (Lawhon, 2009)—and the first residents moved into Radburn, New Jersey—a vanguard town for the motor age (New York Times, 1929). Catering to shifting public preferences and burgeoning automobile adoption, these experiments in a new kind of urbanism for a modern society proved wildly successful by the end of the 1930s. Within a few years they had become the mainstream of American urbanization, exemplified by the 1940s' car-centric, segregated Levittown and its derivatives that proliferated in the wake of World War II (Jackson, 1985).

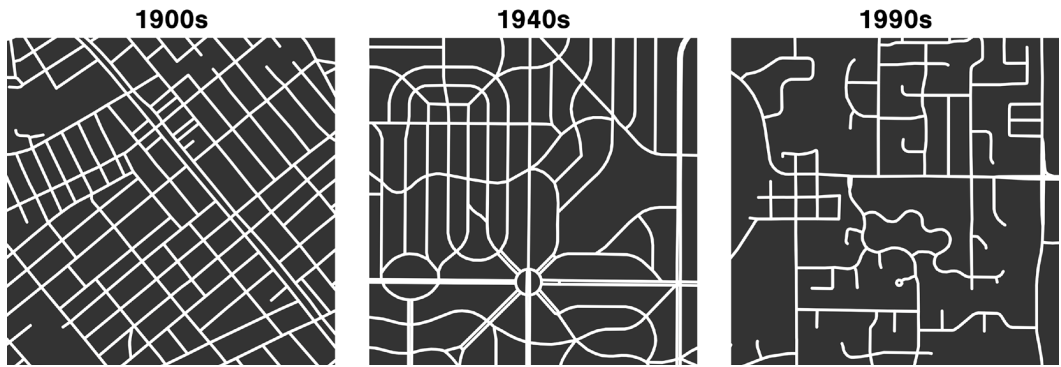

**Figure 2.** Real-world street network patterns from different decades. 1900s: interconnected, imperfect grid. 1940s: increasingly circuitous car-centric suburbs. 1990s: disconnected, dendritic, car-dependent sprawl.

As America suburbanized around the automobile, new technocratic institutions like the Federal Housing Administration (FHA) and Institute of Transportation Engineers (ITE)—both founded in the 1930s—reshaped its spatial form (Southworth and Ben-Joseph, 1995), as illustrated by Figure 2. The FHA underwrote developers' bank loans, but to be deemed a sound investment the proposed subdivision had to adhere to aesthetic standards that—explicitly inspired by Perry and Radburn—embraced automobility and abandoned the grid. Meanwhile, the ITE sought to tame ballooning traffic through geometric design. In 1939, the federal government tasked the ITE with developing its first of many road engineering handbooks. As one such example, ITE's influential 1965 *Recommended Practices for Subdivision Streets* directed planners and engineers to internally-disconnect street networks, avoid four-way intersections, and adopt curving loops and culs-de-sac.

Yet traffic and sprawl swelled unabated, and it was not until the 1990s-2000s that institutionalized street design standards began to emphasize neotradional compactness and connectedness[2]. Southworth and Ben-Joseph (1997) trace these radical overhauls of street network ideology through morphological eras of interconnected grids (c. 1900s), fragmented grids (c. 1930s), warped parallels (c. 1940s-60s), loops and lollipops (c. 1960s-70s), lollipops on a stick (c. 1980s-90s), and turn-of-the-millennium neotraditionalism.

## 2.3. Street Network Design: Values and Impacts

The grid fell out of favor during the 20th century in both theory and practice. Modernist polemics decried it as the fountainhead of urban suffering (Kostof, 1991). Mumford (1961) argued that its monotony annihilated all rapport with the local environment, merely commodifying land for endless expansion without any hierarchical or functional order. Meanwhile, planning practice shifted away from dense, interconnected, gridded street networks in a bid to simultaneously attenuate the automobile's negative externalities (e.g., noise, pollution, streetscape blight, congestion, mortality) in residential communities while still empowering the populace to travel by car because it was fast and convenient (Flink, 1990). But the subsequent development of disconnected, coarse-grained neighborhoods discouraged non-motorized trips, stymied mass transit provision, and exacerbated car dependence and its negative externalities.

  Despite their abandonment during the motor age, street grids have been reappraised in recent decades. Grids lend themselves to navigation and legibility (Lynch, 1960; Sadalla and Montello, 1989), the organization of symbolic, important, and memorable places (Kostof, 1991; Lynch, 1984), platting and extension (Ellickson, 2013; Grant, 2001; Lai and Davies, 2020), efficient transportation (Institute of Transportation Engineers, 2010), comfort and wind mitigation (Kenworthy, 1985), and adaptability to technological change (Jackson, 1985). In conjunction with supportive streetscaping, density, and land-use mix, the grid's interconnectedness supports route choice, access, and the human dynamics of social mixing, activity, and encounter (Alexander, 1965; Forsyth and Southworth, 2008; Groth, 1981; Guo, 2009; Moudon and Untermann, 1991; Jacobs, 1995; Zhu et al., 2020). Grids support active travel by providing pedestrians relatively direct routes across the network, without needing to navigate circuitously around culs-de-sac and disconnected blocks. They also support public transit: buses cannot efficiently route through dead-end dominated neighborhoods and bus routing should involve few turns for operational efficiency and user navigability (Brown and Thompson, 2012).

  Today urban planners work in cities choked with automobile gridlock, face intertwined public health crises from physical inactivity and environmental pollution, and struggle to impede climate change. Physical design matters for several reasons. More-connected street networks—of which grids are the *ne plus ultra*—are associated with lower rates of vehicle ownership and reduced GHG emissions (Barrington-Leigh and Millard-Ball, 2017) as well as increased walkability (Adkins et al., 2017; Hajrasouliha and Yin, 2015). In their classic paper, Cervero and Kockelman (1997) argue that higher proportions of four-way intersections and grid-like patterns are associated with reduced single-occupancy vehicle travel. Ewing and Cervero (2010) identify a relatively large elasticity of vehicle kilometers traveled (VKT) with respect to design metrics such as intersection density and street connectivity (cf. Salon et al., 2012; Stevens, 2017). Street network design also impacts travel behavior and safety (Boer et al., 2007; Braza et al., 2004; Dumbaugh and Li, 2011; Ewing and Handy, 2009; Marshall et al., 2014; Marshall and Garrick, 2010). Yet once street networks are initially built, they remain a semi-permanent city backbone that is difficult to change (Bertaud, 2018; Scheer, 2001; Xie and Levinson, 2009). Their initial planning and design thus lock-in circulation patterns and needs for decades.

   These planning processes exist today within broader sustainability contexts that shape practice (Meerow and Woodruff, 2020). For example, the smart growth movement and the CNU promote compact, connected development for more sustainable, healthy communities (Talen and Knaap, 2003). The LEED-ND certification system—developed in part by CNU—rates neighborhoods on sustainable design, including criteria such as street network patterns, compactness, and connectivity (Ewing et al., 2013; Szibbo, 2016). Although various global, national, and local institutions now call for a return to traditional network patterns as a pillar of urban sustainability, we know little about recent implementation and effects. Accordingly, research examining street network design trends can provide important monitoring and evaluation of planning outcomes.

   Most research on street network evolution uses idealized theoretical models, individual case studies (e.g., Strano et al., 2012; Wheeler, 2003), or small-sample cross-sectional analyses (e.g., Southworth and Owens, 1993; Mohajeri and Gudmundsson, 2014). Less is known comprehensively and empirically about trends in neighborhood-scale street network design across the entire US. Recent work by Barrington-Leigh and Millard-Ball (2015) offers a valuable exception, exploring how urbanized areas' average node degrees (i.e., the number of streets connected to each intersection/dead-end) evolved over time, using Census Bureau TIGER/Line shapefiles and a sample of US counties. They estimate street network vintage in multiple ways, including as residences' median year built in each unit of analysis. The present study builds on this past work by analyzing every census tract across the US, estimating vintage in several new ways, and analyzing several new indicators of street network design and sprawl.

## 3. Methods

Given the literature's theorized importance of griddedness, density, and connectivity on travel behavior and VKT, this study asks the following questions. 1) How have griddedness, density, and connectivity outcomes changed in planning practice during the post-war era of ubiquitous automobility and particularly how have they trended in recent years? 2) Controlling for income, commute length, home and household size, local topography, and street network density and grain, what is the relationship between griddedness and car ownership? To answer these questions, it develops new methods to identify and measure griddedness and estimate urban form vintage algorithmically using open data.

### 3.1. Data Collection

This study examines the street network of each US census tract (*N*=74,133) in terms of its vintage. It uses tracts to capture urban patterns at a roughly neighborhood scale, better reflecting how development occurs piecemeal over time than municipal or metropolitan scales would. Tracts are drawn to be reasonably homogenous and consistent over time and generally follow real-world social and physical boundaries. They thus provide sensible spatial units for examining street network "chunks" and related sociodemographic and built environment characteristics from administrative data.

To construct the tract-level street network models, this study uses OSMnx (Boeing, 2017) to download data from OpenStreetMap, a high-quality worldwide mapping project and geospatial data repository (Basiri et al., 2016; Maier, 2014). These data are then assembled into undirected network models, where *nodes* represent intersections and dead-ends and *edges* represent the street segments that link them (for network modeling details and data repository see Boeing, 2019a). These models collectively comprise approximately 19 million nodes and 24 million edges. OSMnx then attaches elevation to each node and calculates each street's grade (i.e., incline).

**Table 1.** List of variables in the study.

| Theme | Variable | Description/Units |
| --- | --- | --- |
| Griddedness | Grid index | Composite index of orientation order, straightness, and four-way intersection proportion |
|  | Orientation order | Extent to which the streets align in the same directions as each other |
|  | Straightness | How straight are the streets on average |
|  | Four-way intersect proportion | Share of nodes that are four-way street junctions |
| Settlement density/scale | Land area | Land area in 1000s of $km^2$ |
|  | Population density | 1000s of persons per $km^2$ |
|  | Single-family detached home prop | Proportion of housing stock that is single-unit detached |
|  | Median rooms per home | Median number of rooms per housing unit |
| Street network scale/connectivity | Intersection density | Number of street intersections per $km^2$ of land area |
|  | Mean street segment length | Average street segment length between intersections/dead-ends, in meters |
|  | Dead-end proportion | Share of nodes that are dead-ends |
|  | Average node degree | Average number of edges incident to each node in the street network |
| Local topography | Node elevations IQR | Interquartile range of node elevations, in meters above sea level |
|  | Mean street grade | Average absolute incline (rise-over-run) of streets in the network |
| Socioeconomics | Mean household size | Average number of persons per household |
|  | Median household income | Median household income, in 1000s of inflation-adjusted 2018 US dollars |
|  | Mean commute time | Average travel time to work, in minutes |
|  | Vehicles per household | Number of vehicles available per household |

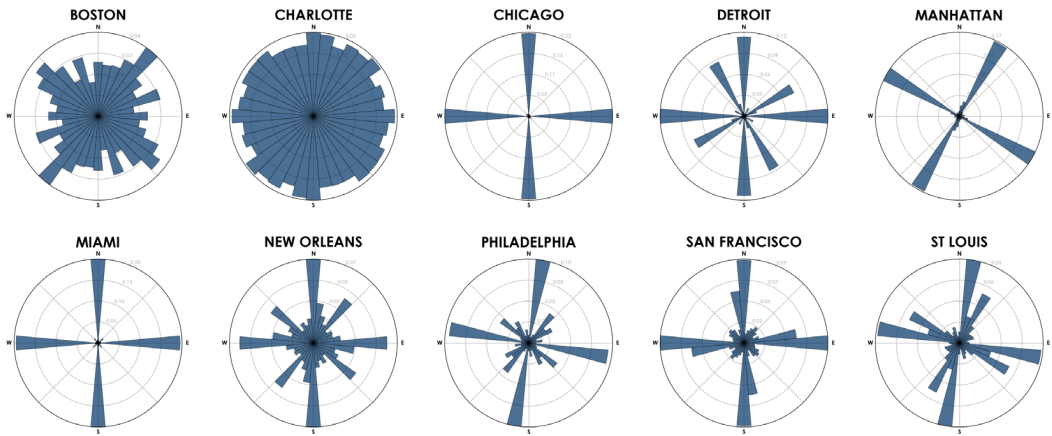

**Figure 3.** Examples of street network orientation for cities proper (except the borough of Manhattan). The polar histogram bars' directions represent compass bearings and bars' lengths represent the proportion of city street segments with those bearings. Chicago and Manhattan illustrate high orientation order (they have internally-consistent orientations as nearly all their streets point in just four different directions), while Charlotte and Boston illustrate low orientation order (their streets point more evenly in all directions). See Boeing (2019b) for methodological details, further interpretation, and worldwide findings.

This study then downloads 2018 US Census Bureau American Community Survey (ACS) tract-level data on built environment characteristics, demographics, and vehicles per household (see Table 1 for variables). Finally, to identify tract vintage, it downloads from the ACS each tract's proportion of residential structures built before 1940 and in each decade[3] since 1940. These variables report the proportion of structures that were first constructed (not remodeled or converted) in each decade, as discussed further below.

Typical quantitative analyses of urban form measure street networks in terms of density, connectivity, and block lengths or areas (Marshall, 2004; Song and Knaap, 2004; Clifton et al., 2008; Song et al., 2013; Porta et al., 2014; Knight and Marshall, 2015; Boeing, 2020; Fleischmann et al., 2020), all of which are operationalized in this study. Once the street network models have been constructed and the census data downloaded, several tract-level indicators are calculated: intersection density, average street segment length, average node degree, node elevation interquartile range (IQR) (a proxy for hilliness; i.e., how much variation exists in a tract's elevation), the proportion of nodes that are four-way intersections, the proportion of nodes that are dead-ends, and whether a tract is urban or not (i.e., population density ≥1,000 persons per square mile, following Census Bureau convention).

### 3.2. Calculating the Grid Index

Next, this study constructs a composite grid index to equally weigh the three components of griddedness identified theoretically in the background section and Figure 1: straightness, orientation order, and the proportion of four-way intersections. Technical details appear in the Appendix, but these components and the index creation process are summarized here.

First, *straightness* is the ratio of the average great-circle distance between each street segment's endpoints and the average length of the street segments themselves. Thus, straightness measures how closely the tracts' streets approximate straight lines.

Second, *orientation order*, developed in detail in Boeing (2019b) and illustrated in Figure 3, measures the relative internal consistency of the streets' orientations. This process calculates the bidirectional compass bearings of every street. It then calculates each tract's street orientation entropy, normalizes it, and linearizes it as an indicator of orientation order. Put simply, this orientation order indicator measures to what extent a tract's streets point in the same directions relative to each other.

Third, the *proportion of four-way intersections* measures what share of a tract's nodes are four-way street junctions. Finally, a composite grid index measuring griddedness is calculated by taking the geometric mean of these three components. Each of the components ranges from 0 to 1. As they are non-substitutable, this uses the geometric mean as a non-compensatory method of aggregation into an index. See the Appendix for technical details on index construction, validation, and robustness.

### 3.3. Tract Vintage Estimation

This study uses ACS structures-built data to examine street network patterns as a function of tract vintage. These data are not perfect: they rely on respondents' memories of construction dates and estimates by longtime residents. They capture information about residential unit construction rather than other built form development, and thus yield proxy estimates of the development era. Nevertheless they provide a useful and best-available approximation of urbanization era nationwide, with a track record in the literature for similar street network vintage estimation and validation (e.g., Barrington-Leigh and Millard-Ball, 2015; Fraser and Chester, 2016). Historical data on street networks are scarce, but due to spatial lock-in and path dependence their patterns tend to remain stable once built (Scheer, 2001). Thus, this study examines snapshots today of tract street networks of various vintage.

This study estimates tract vintage algorithmically via three different methods: a *primary* method plus *earliest* and *assessor* methods as robustness checks. The *primary* vintage estimation method operates as follows. If the majority of a tract's structures were built during a certain decade, it tags the tract as "primarily built" in that decade. If no single decade exceeds 50%, it recursively searches for the earliest decade in which at least 40% of its structures were built, then 30%, and so on until it eventually finds a decade to tag.

As a robustness check, this study separately estimates each tract's *earliest*-built decade by identifying the earliest decade in which at least 20% of its structures were built, conforming to the theory that most tracts' street networks were built around the time that their older buildings were constructed, following Fraser and Chester (2016). Barrington-Leigh and Millard-Ball (2015) similarly use these ACS data to validate street vintage, confirming that the results cohere with a smaller-sample parcel-based identification of construction dates.

As a final *assessor*-based robustness check, this study re-estimates vintage using HISDAC-US property records and assessor data (Leyk and Uhl, 2018). The Appendix contains further details on vintage estimation and validation. As different possible biases could influence each different vintage estimation method, this study inspects decadal trends across all three as a robustness test to see how they diverge or cohere with each other.

**3.4. Regression Modeling**

Next, this study estimates a spatial-lag model (Model I) of tract griddedness as a function of vintage. The model's response variable is the grid index and the predictors of interest are eight variables representing tract vintage decade. The model includes controls for settlement scale, street scale, topography, and county-level fixed effects. This study additionally estimates two spatial-error models (Models II and III) of car ownership as a function of griddedness. The response variable is vehicles per household, which doubly serves as a linear proxy for household VKT because the two are strongly correlated (Barrington-Leigh and Millard-Ball, 2017). Model II's predictor of interest is the grid index and Model III's are the grid index's components: straightness, orientation order, and proportion of four-way intersections. Both models include controls for settlement scale, street scale, topography, median household income, jobs proximity, and county-level fixed effects. All three regression models are estimated on urban tracts only, to focus on *city* planning and design. Complete model specification and estimation details appear in the Appendix.

## 4. Findings and Discussion

**4.1. Nationwide Spatial Trends**

Figure 4 maps each tract in the contiguous US by its grid index value. The Great Plains and Midwest exhibit the most grid-like street networks on average, while New England and Appalachia exhibit the least so. Vermont (grid index=0.13), New Hampshire (0.15), West Virginia (0.16), and Maine (0.17) are the least grid-like by median tract, while South Dakota (0.58), Iowa (0.58), Illinois (0.57), and Nebraska (0.55) are the most grid-like. New England's hilly landscape made it difficult to plan large-scale grids and its development occurred incrementally over centuries around an organic network of country roads and paths. Meanwhile, large-scale Midwest platting and subdivision occurred rapidly across vast swaths of relatively flat land during the heyday of the gridded paradigm.

Accordingly, across the relatively flat Great Plains (North Dakota, South Dakota, Nebraska, Kansas, Oklahoma), griddedness is ubiquitous as both urban and rural tracts have median grid index values of 0.52, demonstrating the influence of the Homestead Act and similar historical orthogonal planning instruments across the region. However, in the Northeastern US (Maine, New Hampshire, Vermont, Massachusetts, Rhode Island, Connecticut, New York, New Jersey, Pennsylvania), griddedness is an exclusively urban phenomenon: this region's urban tracts have a median grid index value (0.49) nearly three-times greater than that of its rural tracts (0.18).

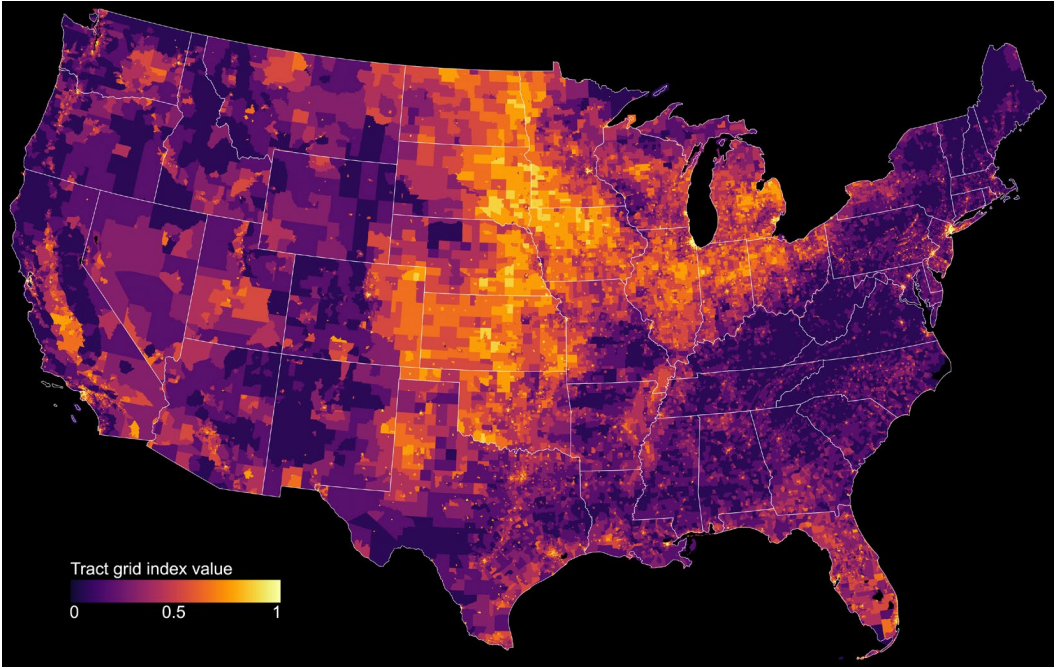

**Figure 4.** Map of tract grid index values across the contiguous US.

This archipelago of urban grids stretches across the Eastern Seaboard and the South in Figure 4. Beyond urban/rural divides, this map also suggests an unsurprising negative relationship between griddedness and mountain ranges such as the Sierra Nevada, Rockies, Appalachians, and Adirondacks. This study isolates these individual relationships between terrain, vintage, urban form, and street network patterns through regression analysis in the following section.

### 4.2. Griddedness and Tract Vintage

The Appendix contains descriptive statistics of the various indicators and decadal averages across all US tracts. The latter represents a snapshot of average indicator values today in tracts of various vintage. The law of constant travel time budgets (Marchetti, 1994) appears to hold as commute times are nearly invariant across vintage, even though street network characteristics and vehicle ownership rates vary substantially between decades.

Controlling for covariates, Table 2 presents the relationship between urban tract vintage and griddedness. Model I has an $R^2$ of 0.74 and its estimated coefficients of interest are all significant. Each decade variable is associated with lower griddedness than the pre-1940 base class. That is, tracts primarily built post-war may be larger, more spread-out, or hillier, but even when controlling for these characteristics, planners and engineers designed these street networks to be less grid-like than was typical prior to 1940. For instance, urban tracts primarily built in the 1980s or 1990s correspond[4] to grid index values 0.15 points lower than those of pre-war tracts, all else equal. Urban tracts primarily built in the 2000s are 0.12 points lower and those built in the 2010s are 0.09 lower.

**Table 2.** Regression model parameter estimates with standard errors in parentheses (significance noted as *p<0.05, **p<0.01, ***p<0.001). County-level fixed effects not shown. Estimated on urban tracts only. Variables' units/descriptions in Table 1.

|  | Grid Index | Vehicles Per Household | |
| --- | --- | --- | --- |
|  | Model I | Model II | Model III |
| Constant | 0.4633*** | 0.8150*** | 0.9869*** |
|  | (0.0421) | (0.1798) | (0.1806) |
| Primarily built 1940s | -0.0352*** | | |
|  | (0.0036) | | |
| Primarily built 1950s | -0.0686*** | | |
|  | (0.0021) | | |
| Primarily built 1960s | -0.1113*** | | |
|  | (0.0025) | | |
| Primarily built 1970s | -0.1340*** | | |
|  | (0.0025) | | |
| Primarily built 1980s | -0.1513*** | | |
|  | (0.0028) | | |
| Primarily built 1990s | -0.1488*** | | |
|  | (0.0030) | | |
| Primarily built 2000s | -0.1184*** | | |
|  | (0.0033) | | |
| Primarily built 2010s | -0.0901*** | | |
|  | (0.0077) | | |
| Grid index | | -0.1809*** | |
|  | | (0.0069) | |
| Straightness | | | -0.2215*** |
|  | | | (0.0239) |
| Orientation order | | | -0.0336*** |
|  | | | (0.0045) |
| 4-way intersection proportion | | | -0.1263*** |
|  | | | (0.0073) |
| Land area | -5.0168*** | | |
|  | (0.2490) | | |
| Population density | 0.0039*** | -0.0055*** | -0.0053*** |
|  | (0.0003) | (0.0003) | (0.0003) |
| Single-family detached home prop | 0.0457*** | 0.4679*** | 0.4716*** |
|  | (0.0036) | (0.0065) | (0.0065) |
| Median rooms per home | -0.0220*** | 0.0269*** | 0.0272*** |
|  | (0.0008) | (0.0018) | (0.0018) |
| Mean household size | | 0.1885*** | 0.1880*** |
|  | | (0.0031) | (0.0031) |
| Median household income | | 0.0036*** | 0.0036*** |
|  | | (0.0001) | (0.0001) |
| Mean commute time | | -0.0031*** | -0.0031*** |
|  | | (0.0002) | (0.0002) |
| Intersection density | 0.0009*** | -0.0006*** | -0.0006*** |
|  | (<0.0001) | (<0.0001) | (<0.0001) |
| Mean street segment length | 0.0003*** | 0.0003*** | 0.0003*** |
|  | (<0.0001) | (<0.0001) | (<0.0001) |
| Node elevations IQR | -0.0012*** | | |
|  | (0.0003) | | |
| Mean street grade | -1.4293** | -0.2339* | -0.2202 |
|  | (0.5353) | (0.1150) | (0.1149) |
| Spatial lag ($\rho$) | 0.3316*** | | |
|  | (0.0101) | | |
| Spatial error ($\lambda$) | | 0.6319*** | 0.6286*** |
|  | | (0.0046) | (0.0046) |
| $n$ | 46208 | 45594 | 45594 |
| Pseudo-$R^2$ | 0.737 | 0.853 | 0.854 |

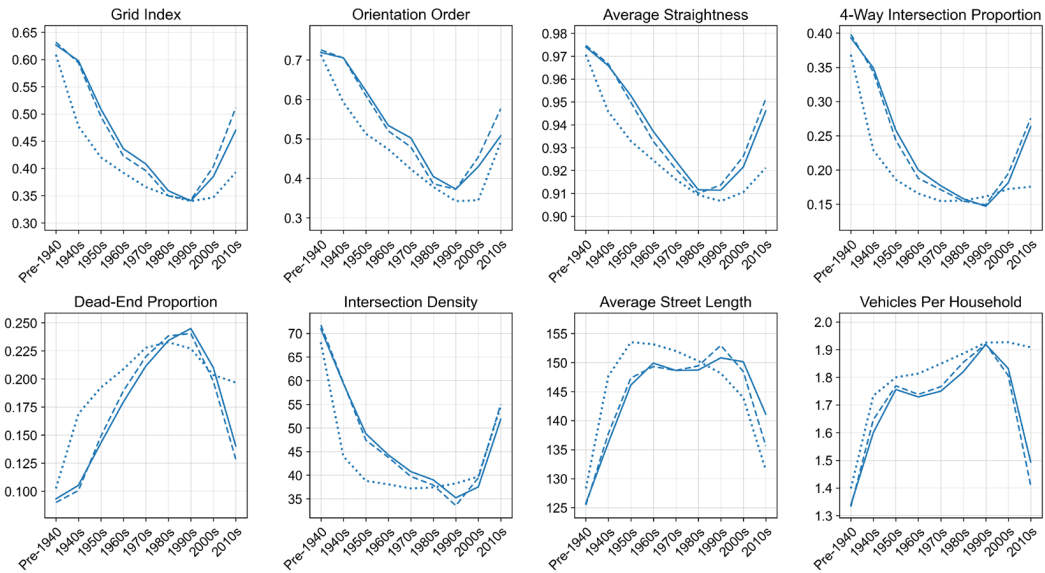

**Figure 5.** Urban tract variables' mean values by vintage. Solid line represents the primary vintage estimation, dashed line represents earliest-decade estimation, and dotted line represents the assessor-based estimation. Variables' units/descriptions in Table 1.

Tract size has a negative relationship with griddedness as larger urban tracts are more likely to comprise amalgams of street orientations and development eras. Greater topographical variation within a tract is associated with less griddedness, suggesting the difficulty of building consistent grids across extreme terrain—though San Francisco provides a well-known exception of engineering a grid irrespective of the underlying landform.

Figure 5 illustrates how key variables in urban tracts trend together across vintage. It presents variables' mean values across all the urban tracts of each decade, showing the primary vintage estimation method alongside the two alternative methods as robustness tests. Compared to the *primary* method, the *assessor*-based robustness test has some important limitations including a non-representative sample that likely underestimates average griddedness while overstating sprawl (see details in the Appendix). Accordingly, the *assessor* trend line tends to peak/trough earlier and demonstrates a more conservative rebound over the past 20 years, bracketing some of these findings. Nevertheless, the indicator values track relatively well across all three estimation methods and their trends across decades tell the same story: griddedness and its constituent components declined steadily from their pre-war highs through the 1990s.

The average grid index value is 84% higher in pre-1940 urban tracts than it is in 1990s-vintage urban tracts, while the four-way intersection proportion is 168% higher. The average proportion of dead-ends is 163% higher in 1990s urban tracts than in pre-1940 urban tracts. Street networks also grew coarser-grained: the average intersection density in pre-1940 urban tracts is double that of 1990s urban tracts, while the 1990s' average street segment length is 20% greater than pre-1940 (equivalent to a 25-meter increase in absolute terms). The average node elevation IQR rose 57% between the 1940s-vintage urban tracts and the 1990s-vintage urban tracts, suggesting that US cities developed on hillier terrain during the latter part of the century, perhaps as they expanded beyond coastal or riparian origins and into surrounding hills.

Most interestingly, however, all of these variables' trends have reversed over the past two decades. Since 2000, the grid index and its components have risen back to levels not seen since the mid-20$^{th}$ century. The average grid index value is 13% higher in 2000s-vintage urban tracts than it is in those of the 1990s. Intersection density is 6.5% higher, the four-way intersection proportion is 23% higher, and the dead-end proportion is 14% lower.

### 4.3. Griddedness and Car Ownership

Figure 5 also reveals that car ownership follows a vintage trend similar to these various street network indicators, rising from 1940 through the 1990s before declining in post-2000 tracts. The average urban tract of pre-1940 vintage has 1.3 vehicles per household today, but the average 1990s tract has 1.9. In other words, households in 1990s-vintage urban tracts own approximately 50% more cars on average than those in pre-war tracts do. However, tract vintage correlates with other important factors like household size, income, and job proximity.

Controlling for such covariates, Model II estimates the relationship between car ownership and griddedness at the urban tract-level with a full set of controls (Table 2). A unit increase in the grid index is associated with a decrease of approximately 0.18 vehicles per household. Re-estimating Model II as a standardized regression reveals that the grid index has the largest effect size[5] among the predictors outside of three socioeconomic variables (household income, household size, and single-family home proportion).

One potential limitation here is that some of the most grid-like neighborhoods are old enough to pre-date the zoning logic of modern functional segregation. Thus, there may be an unobserved factor pertaining to traditional land use patterns influencing car ownership in older gridded neighborhoods. As a robustness test, Model II is re-estimated on only those tracts of 2000s or 2010s vintage to better isolate design in the modern regulatory context. The results ($n$=3618, $R^2$=0.89) remain substantively similar, including an estimated coefficient on the grid index of -0.14 ($p$<0.001). As a final robustness test, Model III decomposes the grid index into its constituent components and predicts vehicles per household using them instead. Each index component is significantly and negatively associated with car ownership. Models II and III have essentially identical $R^2$ values (0.85) and yield similar parameter estimates, suggesting the stability of the index's construction and interpretation, as well as its usefulness as a one-dimensional indicator of griddedness.

### 4.4. The Death and Life of Great American Grids

For over a century, American spatial planning deployed the orthogonal grid as its primary mode of geometric ordering. But new transportation technologies and cultural preferences emerged in the early 20$^{th}$ century to challenge its theoretical and practical prominence. As Radburn's co-chief architect Clarence Stein (1951, p. 41) put it, "The flood of motors had already made the gridiron street pattern, which had formed the framework for urban real estate for over a century, as obsolete as a fortified town wall." Planners, designers, engineers, and developers turned to new network patterns to accommodate the automobile, but accommodation soon grew into dependence. This study takes 1940 as a rupture point—following 20 years of rising car adoption—when massive state intervention in rebuilding mobility infrastructure around the technological and spatial logic of the automobile began to fully dominate American urbanization.

The stark effects of this rupture can be seen throughout this study's findings. Controlling for tract size, terrain, and building and street scale, each decade of vintage after 1939 is associated with lower griddedness than pre-1940 vintage. Griddedness, orientation order, straightness, 4-way intersections, and intersection density all declined steadily from pre-1940 vintage through 1990s vintage, while dead-ends, block lengths, and car ownership rates steadily rose. Urban planners and engineers reorganized cities around the logic of and demand for the automobile after World War II, and we can clearly see this inscribed in the urban form of tracts of different vintage today.

But, importantly, these trends have slowed or reversed since the year 2000, though not to pre-war levels. Nevertheless, post-2000 urban tracts exhibit griddedness, density, and connectivity not seen since the 1950s-1960s as well as lower vehicle ownership rates than any other decade post-1939, though the *assessor*-based method exhibits a more conservative trend in the latter. These findings—across multiple indicators and every urban census tract—are consilient with other research, including Barrington-Leigh and Millard-Ball's (2015) node-degree finding and the historical-morphological case studies of Southworth and Ben-Joseph (1997).

## 5. Conclusions

### 5.1. Summary of Findings

This study developed a new computational big data approach to model the entire US street network per-tract, estimate and compare vintage in different ways, and calculate a basket of new measures and indicators of street network design and sprawl. It modeled the relationship between vintage and form and the relationship between vehicle ownership and griddedness to explore car adoption and dependence in the context of griddedness itself. As is true of all algorithms, this study's vintage estimation methods are imperfect and need not be the final word: in the future, new algorithms should be developed and tested to further home in on precise trends over time. Additionally, the role of urban renewal and redevelopment remains only partly understood. Future work should explore how urban renewal altered preexisting street patterns over time and unpack variation between places in each decade to understand the local contexts that guided these histories heterogeneously within each era.

Keeping these limitations and opportunities in mind, this study made two primary empirical contributions. First, it comprehensively measured how nationwide street network design grew more coarse-grained, disconnected, and circuitous between 1940 and the 1990s and identified a nascent return to classic urbanism patterns over the past 20 years across multiple dimensions. This is a promising though preliminary trend given the impacts of car dependence, VKT, and emissions on urban sustainability, public health, and social justice. Putting numbers to these trends helps monitor and assess planning practice's outcomes and progress toward sustainability goals. Second, it identified new relationships between urban form, vintage, and car dependence while controlling for related characteristics and—importantly—local topography. Previous studies have not fully unpacked these relationships between topography and street network form, or held topography constant to explore other relationships. Beyond these empirical contributions, it developed a novel grid index that offers a new lens to measure and compare urban patterns quantitatively in a theoretically-sound way. Finally, it developed a new set of urban form vintage estimation methods and evaluated their robustness.

## 5.2. Implications for Planning Practice

In recent years, urban planning and public health scholars have identified significant relationships between VKT, road safety, active travel behavior, and street design variables like block lengths, intersection densities, and four-way intersection proportions. The body of theory arising from this research emphasizes the importance in planning practice of network connectivity and density for active travel, safety, and accessibility—yet planners, designers, engineers, and developers steadily drifted away from such connectivity and density as they abandoned the grid and embraced sprawl during the 20th century. This occurred in both greenfield and redevelopment projects: as but one example of the latter, redevelopment efforts in Detroit through the late 20th century typically destroyed the project area's grid to fuse the original blocks into a single superblock (Ryan, 2006).

But there exists another path forward for planning practice. As discussed earlier, new certification standards like LEED-ND promote neotraditional grid-like street network patterns as instruments of sustainability planning in new communities. But even redevelopment and retrofitting projects offer practitioners important opportunities. For instance, Syracuse, NY is currently planning to tear down its 1950s-era Interstate 81 freeway and restore the original street grid (Scheer, 2020). Such planning decisions can help shift cities away from car dependence and back toward finer-grained, interconnected street networks that support active transportation. Meanwhile, in new communities, planners might require minimum intersection densities and proportions of four-way junctions, or use supportive certification standards like LEED-ND to cut red tape. Incentivizing infill development offers further opportunities to take advantage of preexisting urban grids rather than relying on neotraditional greenfield projects at the urban fringe far from job centers.

A shift back toward traditional network patterns may also reflect market demand for more walkable places. As Handy (2017) points out, while academics continue to home in on relevant explanatory variables and regression elasticities, practicing planners, designers, engineers, and developers have recently proceeded with more-compact development for myriad reasons beyond theoretical VKT reductions. Such projects offer clear economic and equity benefits, and can be profitable due to market demand (Kim and Bae, 2020). This is an important arrow in planners' quivers as they advocate for more sustainable street patterns.

In sum, what does this all mean for planning practice? This study's findings can be read as something of a scorecard for the profession over the past 80 years by assessing the trajectory of street network planning and design. Privileging automobility over all other modes, 20$^{th}$ century planners locked-in generations of car dependence. But since the 1990s, planning scholars and prominent practitioners have called for better evidence-based practice to create more sustainable, healthy, and just cities. To successfully implement climate action plans or attenuate pervasive car dependence, practitioners must plan for denser, interconnected networks that allow for non-motorized travel and mass transit provision. This study finds preliminary evidence of some promising trends in this direction toward more sustainable urban forms.

This includes a key takeaway message for practitioners: the initial layout of streets and attendant land parcelization determine urban spatial structure for centuries, locking in mobility needs and capabilities for generations to come. Due to this spatial lock-in, street network patterns are difficult to change once established. So what can practitioners do today?

First, individual suburban retrofits can improve connectivity but are limited by the path dependence of infrastructure and land parcelization. Second, larger redevelopment projects offer strategic opportunities to incorporate (or restore) fine-grained, highly-connected circulation networks into their design. Good design can mitigate historical criticisms of the grid's monotony by providing public space, human-scaled streetscaping, and façade variation. Third, greenfield development may offer practitioners the most straightforward opportunity to continue the aggregate trend back toward more-connected patterns, but such projects are often disconnected from the rest of the urban fabric and far from job centers. Finally, interconnected and relatively fine-grained grids already exist in the cores and inner-ring suburbs of most large US cities. Instead of building new grids on the urban fringe, planners can promote infill and densification where the physical infrastructure already best supports active transportation and freedom of mode choice. Overall, planners and policymakers should review and revise codes and design guidelines at local, state, and federal levels to encourage and streamline the development of networks that support broader sustainability and public health goals.

Interconnected grids defined American spatial patterns for over a century before the rise of the automobile. Across a basket of indicators, this study identified planners' morphological response to and exacerbation of this rise. Rather than merely reacting to fleeting mobility trends, it is imperative for practitioners to plan proactively for the dense, interconnected networks that can attenuate car dependence and advance city climate action plans.

## Acknowledgments

This study was funded in part by a grant from The Public Good Projects. Various pieces of this work in progress were presented over the years at the Association of Collegiate Schools of Planning annual conference, the Urban Affairs Association annual conference, and the Transportation Research Board annual meeting. I wish to thank those conference session participants for their feedback and especially thank Adam Millard-Ball, Lisa Schweitzer, Sara Jensen Carr, Jana Cephas, and Jake Wegmann for their invaluable comments on early drafts.

## Notes

[1] Per the Oxford English Dictionary.
[2] See, for instance, ITE's 1993 Guidelines for Residential Subdivision Street Design, ITE's 1994 Traffic Engineering for Neo-Traditional Neighborhood Design, Oregon's 2001 Neighborhood Street Design Guidelines, and the 2009 LEED-ND Neighborhood Pattern and Design certification criteria.
[3] As these data are from the 2018 ACS, the 2010s decade does not cover the entire decade and thus includes a smaller set of tracts (see Appendix Table A3).
[4] See Appendix for details on interpreting parameter estimates in a spatial-lag model.
[5] That is, beta coefficient magnitude.

Boeing, G., 2020. A Multi-Scale Analysis of 27,000 Urban Street Networks: Every US City, Town, Urbanized Area, and Zillow Neighborhood. Environment and Planning B: Urban Analytics and City Science 47, 590–608. https://doi.org/10.1177/2399808318784595

Boeing, G., 2019a. Street Network Models and Measures for Every U.S. City, County, Urbanized Area, Census Tract, and Zillow-Defined Neighborhood. Urban Science 3, 28. https://doi.org/10.3390/urbansci3010028

Boeing, G., 2019b. Urban Spatial Order: Street Network Orientation, Configuration, and Entropy. Applied Network Science 4, 67. https://doi.org/10.1007/s41109-019-0189-1

Boeing, G., 2017. OSMnx: New Methods for Acquiring, Constructing, Analyzing, and Visualizing Complex Street Networks. Computers, Environment and Urban Systems 65, 126–139. https://doi.org/10.1016/j.compenvurbsys.2017.05.004

Boer, R., Zheng, Y., Overton, A., Ridgeway, G.K., Cohen, D.A., 2007. Neighborhood Design and Walking Trips in Ten U.S. Metropolitan Areas. American Journal of Preventive Medicine 32, 298–304. https://doi.org/10.1016/j.amepre.2006.12.012

Braza, M., Shoemaker, W., Seeley, A., 2004. Neighborhood Design and Rates of Walking and Biking to Elementary School in 34 California Communities. American Journal of Health Promotion 19, 128–136. https://doi.org/10.4278/0890-1171-19.2.128

Brown, J., Thompson, G., 2012. Should Transit Serve the CBD or a Diverse Array of Destinations? A Case Study Comparison of Two Transit Systems. Journal of Public Transportation 15, 1–18. https://doi.org/10.5038/2375-0901.15.1.1

Cervero, R., Kockelman, K., 1997. Travel Demand and the 3 Ds: Density, Diversity, and Design. Transportation Research Part D: Transport and Environment 2, 199–219. https://doi.org/10.1016/S1361-9209(97)00009-6

Clifton, K., Ewing, R., Knaap, G., Song, Y., 2008. Quantitative analysis of urban form: a multidisciplinary review. Journal of Urbanism 1, 17–45. https://doi.org/10.1080/17549170801903496

Dumbaugh, E., Li, W., 2011. Designing for the Safety of Pedestrians, Cyclists, and Motorists in Urban Environments. Journal of the American Planning Association 77, 69–88. https://doi.org/10.1080/01944363.2011.536101

Dunham-Jones, E., Williamson, J., 2011. Retrofitting Suburbia: Urban Design Solutions for Redesigning Suburbs. John Wiley & Sons, Hoboken, NJ.

Ellickson, R.C., 2013. The Law and Economics of Street Layouts: How a Grid Pattern Benefits a Downtown. Alabama Law Review 64, 463–510.

Ewing, R., Cervero, R., 2010. Travel and the Built Environment: A Meta-Analysis. Journal of the American Planning Association 76, 265–294. https://doi.org/10.1080/01944361003766766

Ewing, R., Greenwald, M.J., Zhang, M., Bogaerts, M., Greene, W., 2013. Predicting Transportation Outcomes for LEED Projects. Journal of Planning Education and Research 33, 265–279. https://doi.org/10.1177/0739456X13482978

Ewing, R., Handy, S., 2009. Measuring the Unmeasurable: Urban Design Qualities Related to Walkability. Journal of Urban Design 14, 65–84. https://doi.org/10.1080/13574800802451155

Fleischmann, M., Romice, O., Porta, S., 2020. Measuring Urban Form: Overcoming Terminological Inconsistencies for a Quantitative and Comprehensive Morphologic Analysis of Cities. Environment and Planning B: Urban Analytics and City Science. https://doi.org/10.1177/2399808320910444

Flink, J.J., 1990. The Automobile Age. MIT Press, Cambridge, MA.

Forsyth, A., Southworth, M., 2008. Cities Afoot: Pedestrians, Walkability and Urban Design. Journal of Urban Design 13, 1–3. https://doi.org/10.1080/13574800701816896

Fraser, A., Chester, M.V., 2016. Environmental and Economic Consequences of Permanent Roadway Infrastructure Commitment. Journal of Infrastructure Systems 22, 04015018. https://doi.org/10.1061/(ASCE)IS.1943-555X.0000271

Grant, J., 2001. The Dark Side of the Grid: Power and Urban Design. Planning Perspectives 16, 219–241. https://doi.org/10.1080/02665430152469575
20

# Technical Appendix

### Data Availability

The tracts' street network models are publicly available for others to reuse from the data repository at https://dataverse.harvard.edu/dataverse/osmnx-street-networks. These models are distributed as GraphML files, ESRI shapefiles, and network node/edge lists. Dozens of calculated indicators for each tract are also available at that repository. For more details on the models and repository, see Boeing (2019a). For details on the OSMnx software for modeling and analyzing street networks, see Boeing (2017).

### Grid Index Calculation and Robustness

The grid index is a composite of the three theoretical components of griddedness introduced in the background section: orientation order, straightness, and four-way junctions (see Figure 1). Straightness, $\varsigma$, is calculated as:

$$\varsigma = \frac{D}{L} \tag{1}$$

where $D$ represents the average great-circle distance between each street segment's endpoints and $L$ represents the average length of street segments along the network. Thus, $\varsigma$ measures how closely the streets approximate straight lines.

To calculate orientation order, this study first calculates each tract's orientation entropy, a measure of relative street directionality coherence developed in detail in Boeing (2019b). Entropy measures how ordered or disordered a system or dataset is. This process calculates the bidirectional compass bearing of each street (e.g., a street simultaneously has bearings of both 90° and 270°). Then it calculates the tract's street orientation entropy, $H_O$, as:

$$H_O = -\sum_{i=1}^{n} P(O_i) \, log_e \, P(O_i) \tag{2}$$

where $n$ represents the total number of bins (i.e., 36, such that each bin covers 10 degrees around the compass), $i$ indexes the bins, and P($O_i$) represents the proportion of streets that fall in the $i^{th}$ bin. The maximum entropy, $H_{max}$, that any tract could have equals the logarithm of the number of bins: 3.584. The real-world minimum entropy, $H_G$, is that of a perfect grid and equals 1.386. This normalizes (via min-max scaling) then linearizes street orientation entropy as an indicator of orientation order, $\varphi$:

$$\varphi = 1 - \left(\frac{H_O - H_G}{H_{max} - H_G}\right)^2 \qquad (3)$$

A composite grid index, $G$, can now be constructed measuring how grid-like a tract's street network is by taking the geometric mean of the three components (i.e., the cube-root of their product):

$$G = \sqrt[3]{\varsigma \times \varphi \times \gamma} \qquad (4)$$

where $\gamma$ is the tract's proportion of nodes that are 4-way intersections, and the other variables are as defined above. Each of the components ranges from 0 to 1. As they are non-substitutable, the geometric mean is used as a non-compensatory method of aggregation into an index.

This study calculates this grid index via four different methods: a main method plus three alternative methods as robustness checks. As each index component ranges from 0 to 1, they are effectively min-max scaled, so the main method simply calculates the grid index as described above. As an initial test, the individual components each correlate more strongly with the composite index than they do with one another.

However, the variables' spreads differ, so this study computes alternative specifications as robustness tests to see if this affects the findings' substance and interpretation. The first alternative clips each component's vector of observations to ±3 standard deviations above/below the mean then min-max rescales them before calculating the grid index. While these upper/lower bounds discard information, they attenuate outlier effects and make the vectors' standard deviations more comparable. The second alternative standardizes (i.e., mean-normalizes) each component vector then min-max rescales them before calculating the grid index. The third alternative quantile-transforms each component vector then min-max rescales them before calculating the grid index. This is even more robust against outliers than the other methods.

Finally, this study recomputes all of its descriptive statistics and re-estimates all of its models using each formulation of the grid index. These robustness checks do not substantively alter the findings: the decadal trends and models' interpretations hold across all the various reformulations.

**Vintage Estimation and Validation**

The methods section summarizes the three different tract vintage estimation methods that are operationalized: a *primary* method plus an *earliest* method and an *assessor* method (using HISDAC-US data) as robustness checks. These robustness checks are important because vintage estimates could be noisy or biased due chiefly to three mechanisms.

First, some tracts' streets may have been laid out by visionary planners in one decade, with developers constructing buildings at a later date. New York's 1811 Commissioners' Plan offers one exceptional example. Second, some tracts whose structures were primarily built before 1940 have likely experienced some re-design and re-engineering of their street patterns that reflect later standards. Third, some tracts may have experienced large-scale building turnover via urban renewal, affecting their building-based vintage estimates, without changing their underlying street patterns. Each of these mechanisms could affect each vintage estimation method differently. Fortunately, instances of these three mechanisms are the exception overall, but this study checks for the influence of such possibilities with three diagnostic analyses: vintage validation, robustness tests, and an evaluation of subsequent model estimate precision.

To validate the vintage estimates, dozens of tracts are manually inspected to compare the estimates against historical documentation of construction dates. Appendix Table A1 lists a small sample of these inspected tracts. The influential street network designs of Ladd's Addition, Park La Brea, and Park Merced all have the correct vintage identified across all three methods. The more modern designs of Irvine's El Camino Real neighborhood and Las Vegas's Summerlin West neighborhood are similarly universally correctly identified. Laguna West—one of Southworth and Ben-Joseph's (1995) case studies—and Levittown are correctly identified in both ACS-derived estimation methods. The Chandler tract has slightly inconsistent vintage estimates between the ACS methods (1980s versus 1990s), reflecting its development across different projects in two decades. Such heterogeneity is important to consider as some tracts of course do not neatly build-out entirely within a single decade. The Chandler case demonstrates how the robustness check of multiple estimation methods provides different windows into heterogeneous vintage, which do track with each other overall, as illustrated by Figure 5.

**Appendix Table A1**: A sample of census tracts and their vintage estimates.

| Tract ID | Location | Primary | Earliest | Assessor |
|---|---|---|---|---|
| 25025020101 | Beacon Hill, Boston, MA | Pre-1940 | Pre-1940 | Pre-1940 |
| 41051001102 | Ladd's Addition, Portland, OR | Pre-1940 | Pre-1940 | Pre-1940 |
| 06037214503 | Park La Brea, Los Angeles, CA | 1940s | 1940s | 1940s |
| 06075033204 | Park Merced, San Francisco, CA | 1940s | 1940s | 1940s |
| 36059409200 | Levittown, NY | 1940s | 1940s | Pre-1940 |
| 25025020301 | West End, Boston, MA | 1960s | 1960s | Pre-1940 |
| 06059052526 | El Camino Real, Irvine, CA | 1970s | 1970s | 1970s |
| 04013810700 | Chandler, AZ | 1990s | 1980s | 1990s |
| 06067009619 | Laguna West, Elk Grove, CA | 1990s | 1990s | 2000s |
| 06001403100 | Downtown, Oakland, CA | 2000s | Pre-1940 | Pre-1940 |
| 32003005824 | Summerlin West, Las Vegas, NV | 2000s | 2000s | 2000s |

Similarly, we can see how the vintage estimation methods handle urban renewal. For example, Boston's West End was razed in the 1950s and rebuilt in the early 1960s with a new street network. Both ACS-derived estimation methods correctly tagged its vintage as 1960s, while the assessor data reflect older property records. Boston's adjacent but preserved Beacon Hill neighborhood is correctly tagged as pre-1940 vintage across all methods. Downtown Oakland, meanwhile, experienced a redevelopment boom in the 2000s, which its primary estimate reflects, but its other estimates correctly identify the pre-1940 vintage of its street network. Across the dataset, the law of large numbers appears to hold and the robustness checks generally cohere in Figure 5. Nevertheless, errors can introduce bias and noise into the dataset, underscoring the importance of the robustness checks and the precision of the subsequent model estimates. Regarding the latter, if the data became too noisy from vintage error, it would be impossible to estimate statistically significant parameters in these models. But as we can see in Table 2, that is not the case.

The *assessor* method in particular helps us test against urban renewal effects by using a completely separate vintage data product than the ACS: the HISDAC-US historical settlement dataset (Leyk and Uhl, 2018). These data derive from the Zillow Transaction and Assessment Dataset (ZTRAX), which contains hundreds of millions of property and assessor records across the country. HISDAC-US uses these ZTRAX data to produce a raster file of "first built-up" years, at a five-year temporal resolution and 250-meter spatial resolution, representing when the first built structure appears per pixel in ZTRAX. This study calculates the median HISDAC-US raster cell value per US census tract (the typical tract intersects 44 grid cells) to estimate the "typical" earliest-built date across each tract.

HISDAC-US data are much finer-grained than the ACS, but suffer from two important limitations. First, they exhibit spotty coverage in several states, including Louisiana and much of the Midwest. As these regions are relatively grid-like, we should expect urban indicators derived from HISDAC-US vintage estimates to understate average griddedness (and its corollaries) per-decade, while overstating sprawl. Second, as the raster values represent earliest-built dates, they sometimes predate full (sub)urbanization and street network design. For example, some grid cells' values may represent the presence of an initial homestead or farmhouse in an area that otherwise urbanized decades later (see also Scheer [2001] for the influence of pre-urban farmhouses and roads on subsequent urbanization). However, the fine-grain of HISDAC-US and this study's method of aggregation and median-calculation attenuate this concern somewhat.

Figure 5 shows that relative to the ACS-derived estimates, the HISDAC-US estimated values do appear to exhibit some of the biases discussed above. They tend to peak/trough slightly earlier and demonstrate a more conservative rebound over the past 20 years, bracketing some of this study's findings. Nevertheless, the same basic story is revealed across indicators and decades even when using very different vintage estimation methods, lending further confidence in triangulating these findings.

## Regression Model Specification, Estimation, and Interpretation

This study estimates a linear model (Model I) of tract griddedness by decade of vintage. The response spatially autocorrelates so it specifies Model I as a spatial-lag model to control for these spillover effects:

$$y = \rho W y + \beta_0 + \beta_1 X + \varepsilon \qquad (5)$$

where $y$ is an $n \times 1$ response vector, $W$ is an $n \times n$ queen-contiguity spatial weights matrix, $\rho$ is the spatial autoregressive coefficient to be estimated, $\beta_0$ is the intercept, $X$ is an $n \times k$ matrix of $n$ tract-level observations across $k$ exogenous predictors, $\beta_1$ is a $k \times 1$ vector of coefficients to be estimated, and $\varepsilon$ is an $n \times 1$ vector of errors. The matrix $X$ includes the model's predictors of interest—eight dummy variables representing tract vintage, omitting pre-1940 as the base class—as well as controls for settlement scale (land area, population density, proportion of single-family detached housing, median rooms per home), street scale (intersection density, mean street segment length), topography (node elevations IQR, mean street grade), and county-level spatial fixed effects. This study estimates this model via spatial two-stage least squares using two orders of the spatially-lagged predictors as instruments to account for endogeneity, following Anselin and Rey (2014), and with robust standard errors.

This study additionally estimates linear models (Models II and III) to predict car ownership as a function of griddedness. The response variable is vehicles per household. Diagnostics suggest the error term spatially autocorrelates, so it specifies Models II and III as spatial-error models:

$$y = \beta_0 + \beta_1 X + u \qquad (6)$$
$$u = \lambda W u + \varepsilon \qquad (7)$$

where $y$ is an $n \times 1$ response vector, $X$ is an $n \times k$ matrix of $n$ tract-level observations across $k$ exogenous predictors, $\beta_0$ is the intercept, $\beta_1$ is a $k \times 1$ vector of coefficients to be estimated, $u$ is an $n \times 1$ vector of spatially autocorrelated errors, $W$ is an $n \times n$ queen-contiguity spatial weights matrix, $\lambda$ is the spatial autoregressive coefficient to be estimated, and $\varepsilon$ is an $n \times 1$ vector of uncorrelated errors. The matrix $X$ includes the model's predictor of interest (the grid index in Model II, and its constituent components in Model III) as well as controls for settlement scale (population density, proportion of single-family detached housing, median rooms per home, mean household size), street scale (intersection density, mean street segment length), topography (mean street grade), median household income, jobs proximity (via the proxy of commute time), and county-level spatial fixed effects. This study estimates this model via the generalized method of moments. These models are estimated on urban tracts only, to focus on city planning and design.

Model I's estimated coefficients can be difficult to interpret because of its specification as a spatial-lag model. Due to spatial spillover, each coefficient alone does not represent the marginal effect on the response of a unit increase in the predictor. Instead it represents the *direct effect* (DE): what happens locally if you make a unit change in the predictor only in one tract. But also present are *indirect effects* (IE): local spillovers in each tract from a unit predictor change in other tracts. Following Anselin and Rey (2014), if predictor $x_k$ experiences a simultaneous unit change at all locations, we can calculate the *total effect* (TE = DE + IE) of $x_k$ on the response as:

$$\text{TE} = \frac{\beta_k}{1 - \rho} \tag{8}$$

where $\beta_k$ is its estimated coefficient and $\rho$ is the estimated spatial autoregressive coefficient. Total effects can be calculated accordingly from the information presented in the regression table. Note that in the spatial-error models, we can interpret the estimated coefficients the same way we would in a regular aspatial linear regression. Finally, the pseudo-$R^2$ values report the squared correlation between the observed and predicted values of *y*.

**Appendix Table A2.** Descriptive statistics for the variables across all US tracts (urban + rural) with observations.

|  | count | mean | std dev | min | 25% | 50% | 75% | max |
|---|---|---|---|---|---|---|---|---|
| Grid index | 72659 | 0.421 | 0.209 | 0.000 | 0.256 | 0.396 | 0.561 | 1.000 |
| Orientation order | 72659 | 0.484 | 0.316 | 0.002 | 0.186 | 0.466 | 0.772 | 1.000 |
| Straightness | 72663 | 0.936 | 0.046 | 0.000 | 0.912 | 0.940 | 0.969 | 1.000 |
| 4-way intersection prop | 72663 | 0.210 | 0.172 | 0.000 | 0.092 | 0.157 | 0.272 | 1.000 |
| Dead-end proportion | 72663 | 0.203 | 0.114 | 0.000 | 0.116 | 0.205 | 0.289 | 1.000 |
| Average node degree | 72663 | 2.528 | 0.296 | 0.200 | 2.332 | 2.513 | 2.719 | 3.686 |
| Intersection density | 72660 | 34.666 | 33.318 | 0.000 | 6.088 | 28.296 | 52.189 | 370.754 |
| Mean street segment length | 72663 | 240.322 | 224.423 | 13.773 | 130.482 | 158.987 | 248.747 | 20407.200 |
| Vehicles per household | 71270 | 1.785 | 0.443 | 0.027 | 1.538 | 1.834 | 2.091 | 4.333 |
| Population density | 72660 | 2.087 | 4.594 | 0.000 | 0.125 | 0.873 | 2.103 | 100.631 |
| Single-fam detached home prop | 72225 | 0.621 | 0.267 | 0.000 | 0.467 | 0.684 | 0.826 | 1.000 |
| Median rooms per home | 72104 | 5.588 | 1.142 | 1.300 | 4.900 | 5.500 | 6.200 | 10.000 |
| Mean household size | 72173 | 2.644 | 0.504 | 1.030 | 2.330 | 2.580 | 2.880 | 12.500 |
| Median household income | 72025 | 64.345 | 32.073 | 2.499 | 42.358 | 57.128 | 78.417 | 250.001 |
| Mean commute time | 72144 | 26.275 | 7.238 | 1.000 | 21.200 | 25.500 | 30.600 | 69.100 |
| Node elevations IQR | 72663 | 20.229 | 44.762 | 0.000 | 3.259 | 9.228 | 20.456 | 1513.799 |
| Mean street grade | 72633 | 0.017 | 0.014 | 0.000 | 0.007 | 0.013 | 0.024 | 1.185 |

**Appendix Table A3.** Tract counts and variables' tract-level mean values by vintage, across all US tracts (urban + rural) with observations.

|  | Pre-1940 | 1940s | 1950s | 1960s | 1970s | 1980s | 1990s | 2000s | 2010s |
|---|---|---|---|---|---|---|---|---|---|
| Count | 21151 | 1550 | 9232 | 5645 | 12368 | 6951 | 7800 | 7160 | 368 |
| Grid index | 0.539 | 0.562 | 0.483 | 0.410 | 0.357 | 0.317 | 0.286 | 0.332 | 0.414 |
| Orientation order | 0.617 | 0.661 | 0.585 | 0.493 | 0.421 | 0.344 | 0.299 | 0.359 | 0.448 |
| Straightness | 0.961 | 0.960 | 0.948 | 0.934 | 0.920 | 0.911 | 0.913 | 0.919 | 0.936 |
| 4-way intersection proportion | 0.315 | 0.321 | 0.241 | 0.185 | 0.149 | 0.134 | 0.116 | 0.144 | 0.213 |
| Dead-end proportion | 0.145 | 0.125 | 0.158 | 0.195 | 0.245 | 0.259 | 0.279 | 0.250 | 0.180 |
| Average node degree | 2.657 | 2.688 | 2.622 | 2.508 | 2.436 | 2.411 | 2.384 | 2.468 | 2.585 |
| Intersection density | 48.037 | 52.482 | 42.806 | 36.808 | 25.663 | 26.698 | 18.427 | 21.374 | 33.922 |
| Mean street segment length | 254.531 | 165.656 | 170.966 | 188.238 | 259.383 | 232.266 | 294.315 | 249.168 | 204.406 |
| Vehicles per household | 1.578 | 1.634 | 1.776 | 1.766 | 1.859 | 1.890 | 2.018 | 1.973 | 1.732 |
| Population density | 3.660 | 3.170 | 2.069 | 2.184 | 1.278 | 1.240 | 0.725 | 0.958 | 1.588 |
| Single-fam detached home prop | 0.558 | 0.637 | 0.703 | 0.622 | 0.611 | 0.602 | 0.675 | 0.682 | 0.484 |
| Median rooms per home | 5.366 | 5.181 | 5.580 | 5.560 | 5.524 | 5.649 | 6.008 | 5.978 | 5.305 |
| Mean household size | 2.515 | 2.790 | 2.736 | 2.678 | 2.621 | 2.668 | 2.702 | 2.802 | 2.667 |
| Median household income | 57.311 | 56.113 | 64.981 | 65.368 | 61.138 | 69.624 | 72.099 | 76.416 | 80.641 |
| Mean commute time | 26.142 | 26.380 | 25.800 | 25.859 | 25.331 | 27.172 | 27.078 | 27.585 | 25.937 |
| Node elevations IQR | 21.303 | 10.378 | 12.052 | 13.19 | 25.056 | 21.627 | 24.871 | 21.311 | 13.956 |
| Mean street grade | 0.019 | 0.014 | 0.015 | 0.016 | 0.017 | 0.017 | 0.018 | 0.015 | 0.013 |